\documentclass[11pt]{article}
\usepackage[margin=1in]{geometry}
\usepackage{lmodern,microtype}
\usepackage{amsmath,amssymb,amsthm,booktabs,needspace}
\usepackage[hidelinks]{hyperref}
\hypersetup{pdftitle={Coexistence of Shill-Proof and Non-Shill-Proof Equilibria in a Public Auction},pdfauthor={Benjamin Marsh}}
\newtheorem{proposition}{Proposition}
\title{Coexistence of Shill-Proof and Non-Shill-Proof Equilibria in a Public Auction}
\author{Benjamin Marsh\\[0.35em]
\small Sei Labs and University of Portsmouth\\
\small \href{mailto:ben@seinetwork.io}{\texttt{ben@seinetwork.io}}}
\date{October 2026}

\begin{document}
\maketitle
\begin{abstract}
A single auction can have both a shill-proof equilibrium and a non-shill-proof equilibrium in the model of Komo, Kominers, and Roughgarden. We give an explicit single item example with three bidder positions, publicly observed actions, positive winning prices, and a strictly regular i.i.d.\ value distribution. Real bidders have strict incentives in both equilibria. An initial public action conveys information about whether the final position is real. The resulting belief changes the middle bidder's response and reverses the first shill's strict preference between two actions. We verify both equilibria and compute their expected revenues.
\end{abstract}

\section*{Overview}
Komo, Kominers, and Roughgarden (KKR) observe that a single auction may admit both shill-proof and non-shill-proof equilibria \cite{KKR2025}. This note gives an explicit construction with public actions, strict incentives for real bidders, positive winning prices, and a strictly regular value distribution. Its key feature is that a public bid can convey information about whether a \emph{future} position is real. This information changes the middle bidder's response and reverses the first shill's strict preference between two actions.

\begin{proposition}\label{prop:main}
In KKR's model with independent membership and independent private values, there is a deterministic, public, single-item auction with three potential bidder positions and two pure perfect Bayesian equilibria, $E_0$ and $E_1$. Equilibrium $E_0$ is weakly shill-proof, with equality of the relevant complete contingent strategy profiles, while $E_1$ is not. When both later positions are real, the shill in position~1 strictly prefers action $0$ in $E_0$ and action $L$ in $E_1$. Every prescribed decision by a real bidder is a strict best response. All winning prices are positive, and both equilibria satisfy KKR's individual rationality, ex post monotonicity, and pruning conditions. Expected revenues are
\[
 R_0=2.08645,\qquad R_1=2.15495.
\]
\end{proposition}

\section{The auction}
Each position in $B=\{1,2,3\}$ is independently real with probability $1/2$. Let $R$ denote the set of real positions. The remaining positions are shills. Real bidders know their own value and membership, and observe all past actions. Shills additionally know $R$, but not unrevealed real values. Real utility is $vx-t$. Every shill maximizes total payments from real bidders. Only the winner pays, and shill payments do not count as revenue. Real values are i.i.d.\ with distribution
\[
\begin{array}{c|ccccc}
v&0&2.6&4.5&5.1&8.5\\\hline
f(v)&0.1&0.4&0.1&0.3&0.1
\end{array}
\]
All terminating decimals are exact. The discrete virtual values are
\[
\left(-\frac{117}{5},\frac9{40},\frac{21}{10},\frac{119}{30},\frac{17}{2}\right),
\]
so the distribution is strictly regular.

\Needspace{6\baselineskip}
Position~1 chooses $0,L$, or $H$.
\begin{itemize}
\item Under $H$, position~1 wins immediately and pays $4$.
\item Under $0$, position~2 may buy for $2.5$. If it rejects, position~3 may buy for $2.4$. A second rejection ends without sale.
\item Under $L$, position~2 chooses $0,L$, or $H$, after which position~3 receives the offer below. Acceptance gives position~3 the item at the offered price. Rejection implements the default.
\end{itemize}
\begin{center}
\begin{tabular}{@{}ccl@{}}
\toprule
Position~2's action & Position~3's price & Default on rejection\\
\midrule
$0$ & $3.2$ & Position~1 wins and pays $1.7$\\
$L$ & $3.2$ & Position~2 wins and pays $3$\\
$H$ & $8$ & Position~2 wins and pays $3.5$\\
\bottomrule
\end{tabular}
\end{center}
Every action is public. Each position acts at most once, and there are ten terminal histories.

\section{Two equilibria}
In both equilibria, a real position~1 uses $(0,L,L,L,H)$, with entries ordered by the five values above. After first action $0$, a real position~2 accepts exactly when its value is positive. A real position~3 accepts exactly when its value exceeds its offered price. A position~2 shill rejects after first $0$ and chooses $0$ after first $L$. A position~3 shill rejects every offer. These shill prescriptions apply at every history and membership configuration. Only two prescriptions differ.
\begin{center}
\begin{tabular}{@{}lll@{}}
\toprule
 & $E_0$ & $E_1$\\
\midrule
First shill & Always $0$ & $L$ if $2,3$ are both real, and $0$ otherwise\\
Real position~2 after $L$ & $(0,0,L,L,H)$ & $(0,0,L,H,H)$\\
\bottomrule
\end{tabular}
\end{center}
Thus the only real type that changes its action is position~2 with value $5.1$ after first $L$.

\paragraph{Beliefs.}
Every real information set has positive probability. Beliefs are given by Bayes' rule. Let $q_e$ be a real position~2's probability that position~3 is real after observing first $L$. In $E_0$, only a real first bidder chooses $L$, independently of position~3's membership, so $q_0=1/2$. In $E_1$, a real first bidder chooses $L$ with probability $(1/2)(0.8)=0.4$. Conditional on position~2 being real, a first shill also chooses $L$ when position~3 is real, an event of probability $(1/2)(1/2)=0.25$. Hence
\[
q_0=\frac12,\qquad q_1=\frac{(0.4)(0.5)+0.25}{0.4+0.25}=\frac9{13}.
\]
Conditional on position~3 being real, its value still has distribution $F$. At an information set of a shill, the membership set is known and the prior over real values is conditioned on the preceding actions of real bidders. This conditioning is well defined even after an earlier deviation by a shill, since every sequence of preceding real actions that can arise at such an information set has positive probability.

\section{Verification}
\subsection{Real bidders}
A real position~2 after first $L$ compares
\[
U_0(v,q)=0,\quad U_L(v,q)=(1-q/2)(v-3),\quad
U_H(v,q)=(1-q/10)(v-3.5).
\]
Indeed, a real position~3 accepts prices $3.2$ and $8$ with probabilities $1/2$ and $1/10$, respectively. A shill rejects. The threshold at which $H$ overtakes $L$ is
\[
T(q)=3+\frac{\tfrac12(1-q/10)}{(2/5)q},\qquad
T(q_0)=\frac{43}{8},\qquad T(q_1)=\frac{337}{72}.
\]
Values $0$ and $2.6$ strictly choose $0$. Since
\[
4.5<\frac{337}{72}<5.1<\frac{43}{8}<8.5,
\]
value $4.5$ strictly chooses $L$, value $8.5$ strictly chooses $H$, and value $5.1$ switches from $L$ to $H$. A real position~1 wins after choosing $L$ exactly when position~2 chooses $0$ and position~3 rejects. Each event has probability $3/4$, independently, in both equilibria. Its utilities from $0,L,H$ are therefore
\[
0,\qquad \frac9{16}(v-1.7),\qquad v-4.
\]
The $H/L$ threshold is $487/70$, between $5.1$ and $8.5$. Thus the prescribed first actions are strict. The remaining real decisions are immediate posted price choices. No offered price equals a support value, so they too are strict.

\subsection{Shills}
A position~3 shill earns no revenue from real bidders by accepting. Rejection either collects a real default payment or also earns zero, so rejection is always optimal. After first $0$, a position~2 shill similarly optimizes by rejecting and allowing a possible real sale to position~3. After first $L$, position~2's actions $0$ and $L$ offer position~3 the same price. On rejection, $0$ additionally collects $1.7$ if position~1 is real, whereas $L$ allocates to the shill. Thus $0$ weakly dominates $L$. If position~3 is real, price $3.2$ yields expected payment $1.6$, while price $8$ yields $0.8$, so $H$ is worse than $0$. If position~3 is a shill, $H$ earns zero and $0$ is weakly better. Whenever a shill is indifferent, its prescribed strategy selects the action of the corresponding real bidder with value zero. This tie breaking convention is part of the equilibrium strategy. In particular, after first action $L$, a shill in position~2 chooses $0$ even when position~1 is also a shill and both $0$ and $L$ yield the same revenue. These arguments hold at every history and also establish optimality against coordinated later shill deviations. It remains to check the first shill. Action $H$ earns zero. With the optimal later continuations just established, its other revenues are shown below.
\begin{center}
\begin{tabular}{@{}lrrr@{}}
\toprule
Later real positions & Root $0$, either equilibrium & Root $L$, $E_0$ & Root $L$, $E_1$\\
\midrule
Neither & $0$ & $0$ & $0$\\
Only $3$ & $2.16$ & $1.6$ & $1.6$\\
Only $2$ & $2.25$ & $1.55$ & $1.7$\\
Both $2,3$ & $2.466$ & $2.435$ & $2.69$\\
\bottomrule
\end{tabular}
\end{center}
For the decisive last row,
\begin{align*}
R(0)&=0.9(2.5)+0.1\cdot 0.9(2.4)=2.466,\\
R_0(L)&=0.5(1.6)+0.4[0.5(3.2)+0.5(3)]+0.1[0.1(8)+0.9(3.5)]=2.435,\\
R_1(L)&=0.5(1.6)+0.1[0.5(3.2)+0.5(3)]+0.4[0.1(8)+0.9(3.5)]=2.69.
\end{align*}
The singleton rows follow from $0.9(2.4)=2.16$, $0.9(2.5)=2.25$, $0.5(3.2)=1.6$, and fallback payments $0.4(3)+0.1(3.5)=1.55$ or $0.1(3)+0.4(3.5)=1.7$. Consequently, $0$ is optimal for the first shill in every membership configuration in $E_0$. In $E_1$, it strictly prefers $L$ precisely when both later positions are real, and strictly prefers $0$ when exactly one is real. When neither is real, every action earns zero. Together with the preceding checks by backward induction, these comparisons cover joint shill deviations and prove sequential rationality of both assessments. The strict comparisons that determine the real bidder actions and the first shill's preference reversal have positive slack and therefore persist under small perturbations. Some later shill choices are intentionally tied, as weak shill-proofness permits.

\subsection{Shill-proofness and KKR's conditions}
Write $\sigma(\theta,R)$ for the complete contingent strategy profile, recording shill values as zero. KKR's weak shill-proofness requires, for all recorded value vectors $\theta$ and $S,S'\subseteq\{i\mid\theta_i=0\}$,
\[
\sigma(\theta,B\setminus S)=\sigma(\theta,B\setminus S').
\]
In $E_0$, each shill uses exactly the strategy of a real bidder with value zero in the same position. This includes prescriptions at histories that are not reached. Real strategies depend only on own value and observed history. The required equality therefore holds. In $E_1$, take recorded values $(0,5.1,0)$. If all positions are real, position~1 chooses $0$ and position~2 buys for $2.5$. Replacing position~1 by a shill makes it choose $L$. Position~2 then chooses $H$ and, after position~3 rejects, pays $3.5$. Thus invariance fails. The last row of the revenue table for shills shows that this change is supported by a strict expected gain.

\paragraph{Individual rationality.}
When all positions are real, a first winner pays $1.7$ only with value at least $2.6$, or pays $4$ with value $8.5$. A second winner pays $2.5$, $3$, or $3.5$ only with value at least $2.6$, $4.5$, or $5.1$, respectively. A third winner accepts a price below its value. Losers pay zero.

\paragraph{Ex post monotonicity.}
KKR's condition compares the payment of each fixed bidder before and after one value coordinate rises. If that bidder pays only after the increase, the earlier payment is zero, so only cases in which the same bidder pays at both profiles need checking. If position~1's value rises, its root action moves from $0$ to $L$ to $H$. The possible positive payment changes for a bidder who pays at both profiles are $1.7$ to $4$ for position~1, $2.5$ to either $3$ or $3.5$ for position~2, and $2.4$ to $3.2$ for position~3. A payment may also remain unchanged. If position~2's value rises, then after a root action of $0$ its payment remains $2.5$, while after a root action of $L$ its action moves from $0$ to $L$ to $H$ in either equilibrium. A retained position~1 winner still pays $1.7$, a retained position~2 winner pays $3$ and then $3.5$, and a retained position~3 winner pays $3.2$ and then $8$. If position~1 chose $H$, its payment remains $4$. Finally, raising position~3's value leaves the offered price and default payment unchanged. It either preserves the winner and payment or makes position~3 the winner at the posted price. Thus no positive payment falls in either equilibrium.

\paragraph{Pruning and admissibility.}
All ten terminal histories occur for some value profile in which all positions are real. This holds in each equilibrium because every root action, every position~2 menu action, and both responses to every posted offer occur. Every decision node has at least two successors. Holding other values fixed, varying the acting bidder's value can change the outcome. For the first two positions, fixing later values at zero witnesses this. The checks above allow every action at every shill information set and therefore establish optimality against all joint shill strategies. This is stronger than KKR's Definition~A.1 requirement to check only the restricted deviation set $\Sigma_S$. Together with the preceding observations, this verifies Definition~A.1 and the pruning conditions in Definition~A.2 \cite{KKR2025} and completes the equilibrium verification.

\Needspace{10\baselineskip}
\section{Revenue and interpretation}
Enumerating the eight equally likely real membership sets gives the following table.
\begin{center}
\small
\begin{tabular}{@{}lrrrrrrrr@{}}
\toprule
Real set & $\varnothing$ & $\{1\}$ & $\{2\}$ & $\{3\}$ & $\{1,2\}$ & $\{1,3\}$ & $\{2,3\}$ & $\{1,2,3\}$\\
\midrule
$E_0$ & $0$ & $1.76$ & $2.25$ & $2.16$ & $2.545$ & $2.576$ & $2.466$ & $2.9346$\\
$E_1$ & $0$ & $1.76$ & $2.25$ & $2.16$ & $2.665$ & $2.576$ & $2.69$ & $3.1386$\\
\bottomrule
\end{tabular}
\end{center}
For a real set $\{1\}\cup S$, with $S\subseteq\{2,3\}$, let $r_e^a(S)$ denote the entry in the preceding revenue table for shills that corresponds to initial action $a$ and later real set $S$. Conditional revenue equals
\[
0.1r_e^0(S)+0.8\bigl[r_e^L(S)+1.7\,2^{-|S|}\bigr]+0.4.
\]
The first bidder's action probabilities are $0.1,0.8,0.1$. Conditional on $L$, its default payment occurs with probability $2^{-|S|}$. Averaging the eight membership cases gives the revenues asserted in Proposition~\ref{prop:main}.
\[
R_0=\frac{41729}{20000},\quad R_1=\frac{43099}{20000},\quad
\frac{R_1}{R_0}=\frac{43099}{41729}\approx1.03283.
\]

The two equilibria sustain different beliefs about future membership. Those beliefs change the middle bidder's action, which in turn changes the early shill's incentive. Because real bidding is permitted to differ across equilibria, the difference $R_1-R_0$ is not itself a measure of the profitability of shilling. Within $E_0$, no shill has a profitable deviation. The example is asymmetric and neither efficient nor optimal for revenue. For instance, when all positions are real and values are $(5.1,4.5,0)$, the second position wins despite having the lower value. Coexistence is also known in other shilling models, including Lamy's model with participation and costs of shill entry \cite{Lamy2008}. The contribution here is an explicit public example under KKR's information structure and membership assumptions.

\paragraph{Acknowledgments.}
I thank Andrew Komo for helpful correspondence.


\begin{thebibliography}{9}
\raggedright
\bibitem{KKR2025}
Andrew Komo, Scott Duke Kominers, and Tim Roughgarden.
\newblock \emph{Shill-Proof Auctions}.
\newblock arXiv:2404.00475, version~3, December~31, 2025.
\newblock \url{https://arxiv.org/abs/2404.00475v3}.

\bibitem{Lamy2008}
Laurent Lamy.
\newblock \emph{Competition between auction houses: a shill bidding perspective}.
\newblock Working paper, January~24, 2008.
\newblock \url{https://www.columbia.edu/~pj2133/ShillFeesParticipationCosts.pdf}.
\end{thebibliography}
\end{document}